# Brain Tumor Detection and Classification with Feed Forward Back-Prop Neural Network

Neha Rani
Student
Electronics and Communication
Engineering
The Northcap University
Gurgaon

Sharda Vashisth, PhD
Associate Professor
Electronics and Communication
Engineering
The Northcap University
Gurgaon

## ABSTRACT
Brain is an organ that controls activities of all the parts of the body. Recognition of automated brain tumor in Magnetic resonance imaging (MRI) is a difficult task due to complexity of size and location variability. This automatic method detects all the type of cancer present in the body. Previous methods for tumor are time consuming and less accurate. In the present work, statistical analysis morphological and thresholding techniques are used to process the images obtained by MRI. Feed-forward back-prop neural network is used to classify the performance of tumors part of the image. This method results high accuracy and less iterations detection which further reduces the consumption time.

## 1. INTRODUCTION
Brain is an organ that controls activities of all the part of the body. Growth of abnormal cells of brain leads to brain tumor. Diagnosis of Brain tumor is very important now-a-days. Tumor basically refers to uncontrolled multiplication of cells. A cell rapidly divided from a micro calcification, Lump, distortion referred to a tumor. Metastasis is a process in which tumor occurring cells moves the other part of the body and tumors begin from that regular tissue reinstate. Meningioma and glioma are the types of brain tumor. Brain tumor is more curable and treatable if detected at early stage; it can increase the intracranial pressure which can spoil the brain permanently. Brain tumor symptoms depend upon the size of tumor, location and its type. Detection of tumor can be done by MRI and CT scan. Brain angiogram procedure can be applied in which blood vessels are illuminated in the brain and feed blood the tumor part. Procedure of biopsy is also including tissues or sample of cells are taken from the brain at the time of surgical treatment, this will help to predict the benign of cancerous brain tumor. Sometimes cancer diagnosis can be delayed or missed because of some symptoms. The principle aim of this paper is to analyze the best segmented method and classify them for a better performance.

## 2. RELATED WORK
MRI provides high qualities of images and visualizes structure of the body internally. Different types of tissues in the body can be distinguished completely with MRI and also contains fine information for treatment (Al-Badarneh et al., 2013; Amasaveni, 2013). Texture of MRI contains information of size, shape, color and brightness that texture properties helps to detect texture extraction (Avula, 2014; Chan et al., 2014). Neural Network (NNs) consists of an interconnected components, it contains the mimic properties of biological neurons. In (Feed-Forward backprop) more than one neuron can be simply defined as interconnected components having large inputs activation function and output (Chanet et al., 2011; Chudler, 2011). The remainder part of this method is organized as follows. Section 2 discusses about related work. Section 3 discusses methodology. Section 4 observes the experimental results of our methodology more than 200 images from MRI dataset and collected data from hospitals.

It is [1,2] proposed that histogram equalization image segmentation and then extracting the feature using Gray level Co-occurrence matrix for detection of tumor. MRI dataset of 120 images are implemented which was made available by Radiology Department of Tats memorial hospital. This method only detects location and size of tumor. In [3,4], this method Bias correction for preprocessing feature extraction and ada-boost classifier for selected features and trained classifier. Dataset of MRI 100 images are gathered from the Hospital of sick children Toronto. Accuracy obtain by this method is 90.11%. In [5,6] the proposed method consists of multiple phases in first page texture extraction of feature. In second phase brain images classify the bases of this feature using RBFN (Radial Basis neural Network) and BPN (Back-prop Neural Network) classifier malignant tumor is segmented by the process of segmentation. In this study [7,8] method consist of a four stages preprocessing image extraction feature testing Rough set Theory (Binary Classifier) and Feed forward Neural Network. 20 MRI brain tumor images are collected from real resources. Classification of MRI images is done at different pathological conditions. In study of [9,10] preprocessing of tumor images train data Dimensionality reduction using DCT and Gabor filter and testing using PNN (Probabilistic neural network). Classifier for training images with accuracy 89.9%.

## 3. METHODS AND MATERIAL
MRI dataset is collected from the Harvard medical school architecture and some data are gathered from civil hospital of Haryana. One of 220 MRI images of normal and abnormal images shown in Fig.1 which is trained with this method.

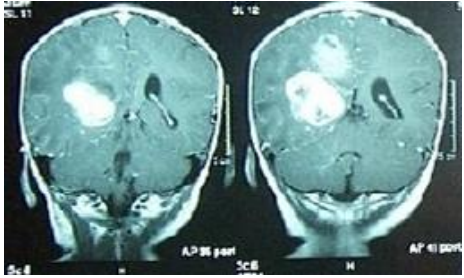

**Fig. 1 Abnormal Training Images**

The proposed method takes the input MRI images that will undergo grey image conversion, template creation, computation of correlation undergoes tumor location detection. Brain tumor segmentation and training. The proposed method takes the input MRI images that will undergo grey image conversion, template creation, computation of correlation undergoes tumor location detection. Brain tumor segmentation and training. The proposed method framework is discussed in Fig.2.

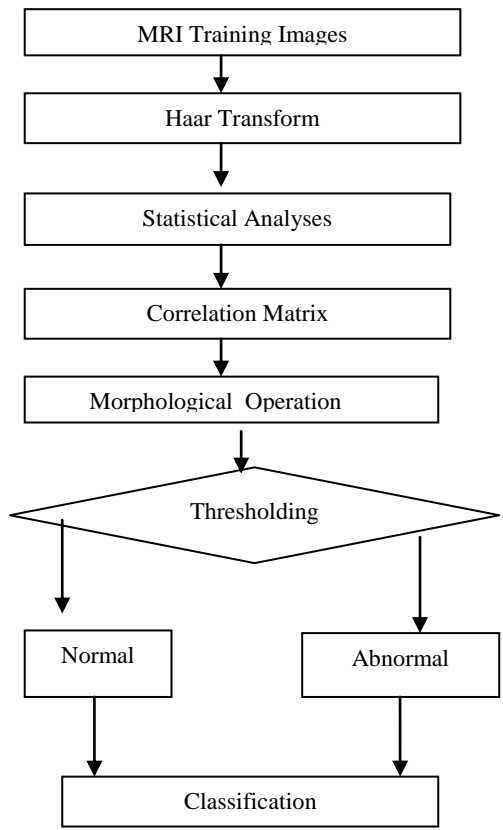

**Fig. 2 Automatic detection workflow**

Filtering is done to remove non-brain tissue. Haar wavelet transform is used for the pre-processing of image. Wavelet coding is suitable for the applications where tolerable degradation and scalability are important. Haar wavelet transform decomposes the input signals into a set of the basis function are called wavelets. A prototype wavelet is called mother wavelet other wavelets are obtain this by shifting or dilations called daughter wavelet.

$$\Psi a, b(t) = \frac{1}{\sqrt{a}} \Psi\left(\frac{t-b}{a}\right) \quad (1)$$

Where a is scaling and b is the shifting parameter. Ψ represents a wavelet function. If the input series is X = ($x_1, x_2......x_N$) haar wavelet transform convert it into one low pass wavelet coefficient series ($L_j$) and one high pass wavelet ($H_j$) series each of length n/2 represented by:

$L_j = \sum_{p=0}^{k-1} X_{2i-p} . T_P(Z)$ ……………..(2)

$H_j = \sum_{p=0}^{l-1} X_{2i-p} . S_p(Z)$ ........................(3)

Where $T_p(Z)$ and $S_P(Z)$ represent wavelet filters. K is filter length and j=0,1,2….[n/2]-1. Statistical property of filtering images is predicted. Basically used to describe the parameter for a specific purpose. Statistical property is used to estimate the distribution parameters. Important property of statistics that includes mean, variance and entropy. The correlation coefficient array is mainly used to detect the target in MRI brain images. It predicts the dependency between the multiple variables at same time. Correlation coefficient matrix detects the part of high intensity area where tumor is present. Identify the relevant feature results is the easier, faster and better understanding of images. Relevant information of input data can be predicted by feature extraction. Algorithms are used to isolate and detect the shapes and desired portions significantly. The quality of the process of feature extracting affects the classification process. Thersholding helps to obtain the normal or abnormal images. Most effective techniques are used to isolate the object by converting in binary image from grey level and image with high contrast levels. Area of extracted image is calculated by summation of black pixels ($B_p$) and white pixels ($W_p$).

IMAGE (I) = $\sum_{WP=0}^{255} . \sum_{Bp}^{255} . [f(0) + f(1)]$……….(5)

f (0) = Black pixels (digit 1)

f (1) = White pixels (digit 0)

Total number of White Pixels = $\sum_{Wp=0}^{255} . \sum_{Bp}^{255} [f(1)]$…..(6)

$P_v$ = Total number of white pixels ($W_p * B_p$)……..(7)

1 Pixel = 0.264583 mm

Tumor Size = [($P_v$) * 0.264583]……………(8)

Classification refers to computational method for data and also finds patterns. Training of tumors images is done with feed forward back-prop neural network (FFBN) by setting various properties. Neural Network consists of neurons, simple components are interconnection to each-other and similar the property of biological neurons. The output obtained from neurons as a function of F(Y) input vector (X $X_2......X_n$).

$Yi = \sum_{j=1}^{k} W_{ij} Xi$…………………..(9)

A multilayer perception network is consists of corresponding neurons and Weighted sum ($W_{ij}$).

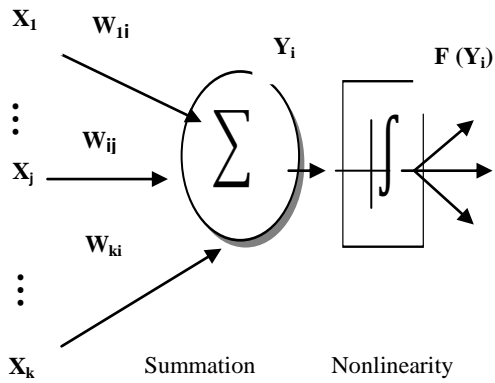

**Fig. 3 Neural network with weighted function and bias function**

MATLAB is used to design and analyze the system. MATLAB is used for signal processing, digital image procession, computer vision, machine learning, control design and communication etc. In MATLAB classification performance is analyzed. MATLAB allows the plotting of data and function, matrix manipulation, implementations of algorithms, matrix manipulations, and interfaces with various programming language.

In FFBP information flow in one direction along all the connecting paths. Information passes through from input to output via hidden layer without loops (feedback). Back propagation is the simplest network to calculate the performance of derivative with respect to weight and each bias variable is adjusted according to the gradient descent. $L_r$ represents the learning rate if performance decreases then learning rate increases. Dx represents the derivative performance w.r.t. variable x. given below.

$$Dx = lr * \frac{dperf}{dx} \ldots\ldots\ldots\ldots\ldots(10)$$

In each iteration training images are reweighted. Evaluation criterion of neural network is to minimize the MSE (Mean Square Error). Basically used to evaluate the performance of the model.

Mean Square Error $= \frac{1}{n}\sum_{j=1}^{n}(Observed - Predicted)^2 \ldots (11)$

Validation, training and testing are used to analyze the performance of the neural network. Learngd is the adaption learning and Traingd is the Training function for multi layer perceptrons in the network. For learning patterns of data in neural network training set, evaluate the generalized ability of trained network testing set and validation set for checking the performance.

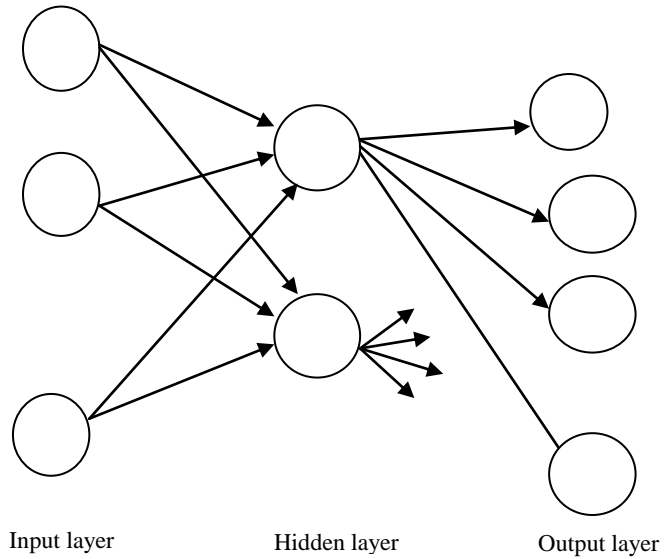

**Fig. 4 Multilayer FFBN**

For Learning function to solve a specific task consider a function F for a class. Learning referred to the find a function F*$\epsilon$ f which helps to solve optimal task. Price function is given below.

$$P : f \to R \ldots\ldots\ldots\ldots\ldots..(12)$$

This is a optimum solution.

$$F * , P\,(F *) \leq P\,(F) ¥ F\, \epsilon\, f \ \ldots\ldots(13)$$

For training Traingd is the training function that updates the weight and bias value according to gradient descent. For every slow iteration algorithm training status displayed. If training status shown by Nan then it means that training status will be never displayed the performance function drops below the goal when number of iteration go high to the epochs. Some training parameters associated with the neural network are epoch, Iteration, goal, time, max_fall, max_fail.

## 4. RESULTS AND ANALYSIS

The result of tumor detection and classification is shown in Fig. 5

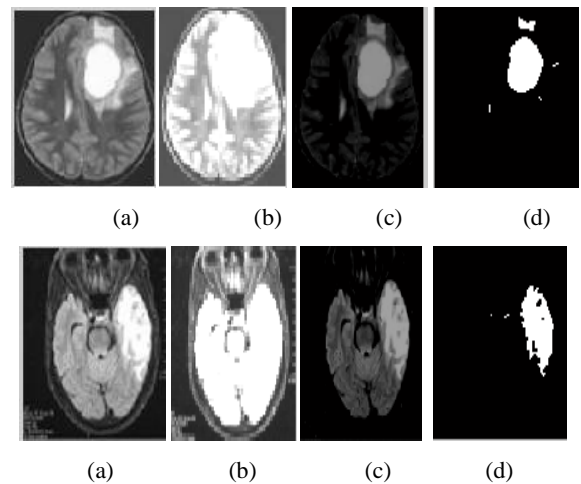

(a)     (b)     (c)     (d)

(a)     (b)     (c)     (d)

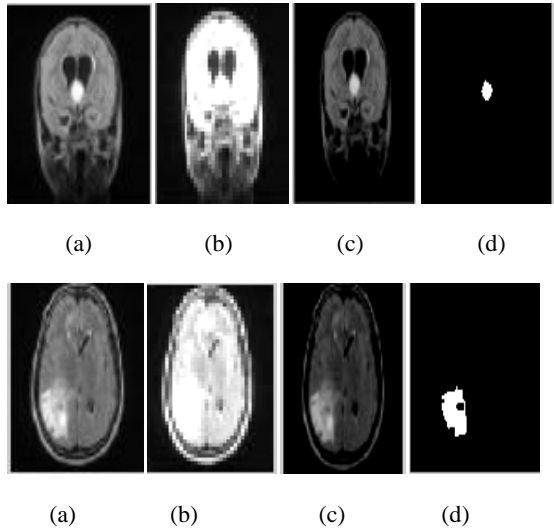

**Fig. 5 (a) MRI Training Image, (b) Filtering, (c) Statistical analyses (d) Segmented Part**

Features of 220 abnormal images are obtained. The features are passed through the feed forward back-prop neural network. The performance of proposed method in term of correlation matrix, entropy, and connectivity, number of objects and number of iteration are evaluated.

**Table 1: Result of MRI Images of different Statistical property**

| Image | Mean | Std. | Variance | Entropy | Objects |
|---|---|---|---|---|---|
| Image1 | 100 | 10.89 | 118.2 | 0.531 | 6 |
| Image2 | 131.30 | 11.87 | 141 | 0.0076 | 38 |
| Image3 | 126.95 | 11.56 | 133.6 | 0.165 | 48 |
| Image 4 | 145.73 | 11.85 | 90.6 | 0.946 | 34 |
| Image5 | 81.29 | 11.58 | 134.09 | 0.7323 | 34 |
| Image 6 | 145.73 | 9.68 | 134.05 | 0.098 | 40 |
| Image7 | 81.29 | 9.97 | 99.42 | 0.067 | 35 |
| Image8 | 145.70 | 9.51 | 134.01 | 0.876 | 13 |
| Image 9 | 70.12 | 8.43 | 127.5 | 0 | 4 |
| Image10 | 145.3 | 12.56 | 157.92 | 0.534 | 6 |
| Image11 | 154.9 | 12.65 | 160.6 | 0.761 | 2 |
| Image12 | 94.72 | 10.98 | 102.63 | 0.6761 | 25 |
| Image13 | 106.94 | 9.66 | 93.36 | 0.045 | 6 |
| Image14 | 60.46 | 10.82 | 117.24 | 0.08 | 11 |

**Table 2: Result of MRI Images of different Statistical property**

| Image | Connectivity | Area(mm$^2$) | Observation | Prediction |
|---|---|---|---|---|
| Image1 | 8 | 19.383 | 2D | Cancer |
| Image2 | 8 | 11.42 | 2D | Cancer |
| Image3 | 8 | 6.056 | 2D | Cancer |
| Image 4 | 8 | 16.887 | 2D | Cancer |
| Image5 | 8 | 13.61 | 2D | Cancer |
| Image 6 | 8 | 1.451 | 2D | Cancer |
| Image7 | 8 | 13.47 | 2D | Cancer |
| Image8 | 8 | 20.34 | 2D | Cancer |
| Image 9 | 8 | 5.772 | 2D | Cancer |
| Image10 | 8 | 12.23 | 2D | Cancer |
| Image11 | 8 | 11.318 | 2D | Cancer |
| Image12 | 8 | 13.223 | 2D | Cancer |
| Image13 | 8 | 12.32 | 2D | Cancer |
| Image14 | 8 | 0.147 | 2D | Cancer |

Properties of feature extraction i.e. mean, standard deviation, variance, entropy, connectivity and number of the objects are obtained. Central tendency value is set if mean, std. and variance value is less or more shows the types of the tumor. If value is less than central tendency that's shows that primary tumor other case secondary tumor. Connectivity shows the image type as 2-D or 3-D. Number of objects shows the degree of tumor spread in the part of the body. Larger the number of objects means presence of more number of tumors in that part. After feature extraction, area of the tumorous region is calculated from which size of the tumor is analyzed.

**Table 3: Result of Classifier at different cases**

| Image | Iteration | Time | Performance | Gradient |
|---|---|---|---|---|
| Image1 | 19 | 1s | 3.72E+03 | 190 |
| Image2 | 18 | 0s | 1.05E+03 | 400 |
| Image3 | 30 | 1s | 5.80E+03 | 387 |
| Image4 | 31 | 1s | 1.86E+03 | 345 |
| Image5 | 18 | 0s | 2.42E+03 | 244 |
| Image6 | 30 | 1s | 6.84E+03 | 49.2 |
| Image7 | 28 | 1s | 1.47E+03 | 116 |
| Image8 | 12 | 4s | 7.37 | 6.5 |
| Image9 | 26 | 0s | 1.21E+03 | 8 |
| Image10 | 6 | 0s | 181 | 22.7 |
| Image11 | 87 | 2s | 1.11E+03 | 7.32 |
| Image12 | 59 | 1s | 1.55E+02 | 13.2 |
| Image13 | 23 | 4s | 1.44E+02 | 62.5 |
| Image14 | 65 | 1s | 1.48E+02 | 40.3 |

**Table 4: Result of Classifier at different cases**

| Image | Validity | Best Performance | Regression |
|---|---|---|---|
| Image1 | 6 | 46.35 | 0.33043 |
| Image2 | 6 | 7369.39 | 0.34239 |
| Image3 | 6 | 715.76 | 0.20913 |
| Image4 | 6 | 6027.98 | 0.64031 |
| Image5 | 6 | 2381.17 | 0.89899 |
| Image6 | 6 | 4099.66 | 0.6255 |
| Image7 | 6 | 2192.24 | 0.1489 |
| Image8 | 6 | 1915.86 | 0.9556 |
| Image9 | 6 | 2788.142 | 0.18322 |
| Image10 | 6 | 934.4 | 0.00962 |
| Image11 | 6 | 1876.15 | 0.23656 |
| Image12 | 6 | 1673.92 | 0.062291 |
| Image13 | 6 | 86.73 | 0.1847 |
| Image14 | 6 | 1207.1 | 0.072463 |

The test and validation curves are very similar. Better performance is indicated if the test curve shows significant increase as compared to validation curve.

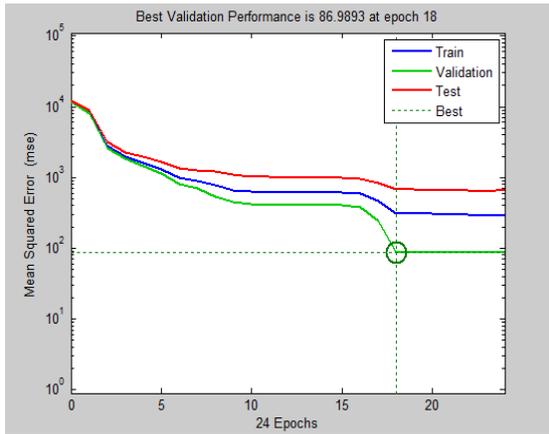

**Fig.6 Neural network Training Performance**

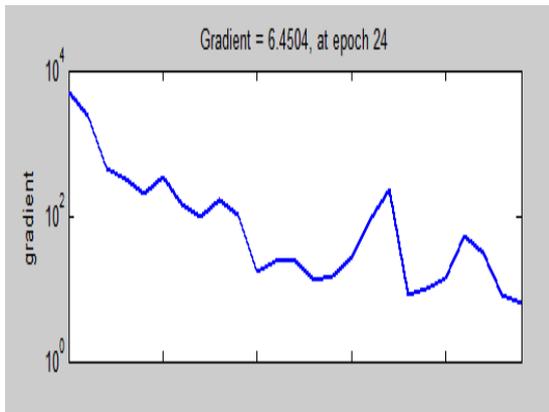

**Fig.7 Training Gradient Performance**

Gradient descent is multiplied with negative descent that shows changes occur in biases and weights. The algorithm becomes stable if the learning rate is small.

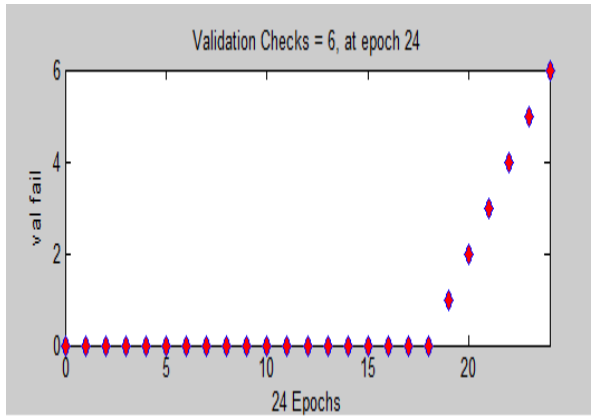

**Fig. 8 Training validation Performance**

Validation performance shows the relation between the output and the target. Maximum validation shows the perfect training of the target. Sensitivity is the ratio of true positive value which are correctly measured by the experimental test, Specificity is the true negative value measured by the experimental test. Accuracy shows that diagnostic test is closer to the true value. This proposed method shows that sensitivity is 98.5 %, specificity is 97.2 % and accuracy is 99.2% shown in Table 3 that is better than existing methods. Table 4 shows the comparison with the literature survey given in paper this method obtain high accuracy.

$$\text{SENSTIVITY} = \frac{(TRUE\ POSITIVE)}{(TRUE\ POSITIVE + FALSE\ POSITIVE)} \ldots\ldots\ldots(14)$$

$$\text{SPECIFICITY} = \frac{(TRUE\ NEGATIVE)}{(TRUE\ FALSE + FALSE\ POSITIVE)} \ldots\ldots(15)$$

$$\text{ACCURACY} = \frac{(TRUE\ NEGATIVE + TRUE\ POSITIVE)}{(TRUE\ NEGATIVE + TRUE\ POSITIVE + FALSE\ NEGATIVE + TRUE\ POSITIVE)} .(16)$$

**Table 5: Experimental Results**

| Evaluation | Proposed Result |
|---|---|
| SENSTIVITY | 98.5 % |
| SPECIFICITY | 97.2 % |
| ACCURACY | 99.2 % |

**Table 6: Shows the Comparisons of Various Results**

| Classification Results | Accuracy |
|---|---|
| Sahar , G. (2012) | 90.11 % |
| Amasaven, V. (2013) | 89.9 % |
| Hussein ,W. (2013) | 96.33 % |
| Machhale, K. (2015) | 98 % |
| Surugavalli, S. (2016) | 96.6 % |
| Proposed Method (2016) | 99.2 % |

## 5. DISCUSSION AND CONCLUSION

This paper shows that combination of feature extraction and classification analysis. After analyzing the results it is concluded that this method is better than the other existing methods in terms of computation time. This automatic segmentation algorithm gives shape and size of the tumor more accurately and other properties like connectivity and the number of objects. Information of images can be obtained by principle component analysis, where the possibility of tumor is highest by using mean, entropy and correlation matrix. Result of the classifier reduces number of iterations and thus the computation time. Validation performance reached maximum. Specificity is 97.2%, Sensitivity is 97.2% and accuracy is 99.2%. Comparison results of proposed methodology with other authors results shows that this method gives more accurate results with the accuracy of 99.2%.Shows that Classifier selection further may be researched to find the better results and This method will be implemented on the 3D images.